\documentclass[printer]{aa}  
\usepackage{graphics,epsf,epsfig,graphicx}  
\begin{document}  
\setlength{\textfloatsep}{0.4cm}
  
\title{ New multi-zoom method for N-body simulations:   
application to galaxy growth by accretion.}

\author{B. Semelin \inst{1} \and  F. Combes \inst{1}}
\offprints{B. Semelin, \email{ benoit.semelin@obspm.fr}}
\institute{
Observatoire de Paris, LERMA, 61 Av. de l'Observatoire,
F-75014, Paris, France
}
\date{}
\authorrunning{Semelin \& Combes}
\titlerunning{Galaxy growth by accretion}

\abstract{The growth of galaxies is driven by two processes: mergers with  
other galaxies and smooth accretion of intergalactic gas. The relative share  
of this two processes depends on the environment (rich cluster or field),  
and determines the morphological evolution of the galaxy. In this work we focus  
on the properties of accretion onto galaxies. Through numerical simulations we  
investigate the geometrical properties of accretion. To span the scale range  
required in these simulations we have developed a new numerical technique:  
the {\sl multi-zoom method}. We run a series of Tree-SPH simulations in smaller   
and smaller boxes at higher and higher mass resolution, using data recorded  
at the previous level to account for the matter inflow and the tidal field from  
outside matter. The code is parallelized using OpenMP. We present a validation test   
to evaluate the robustness of the method: the pancake collapse.  
  
We apply this new  {\sl multizoom method } to study the accretion properties.  
Zooming in onto galaxies from a cosmological simulation, we select a sample of 10   
well resolved galaxies (5000 baryonic particles or more). We sum up their basic  
properties and plot a Tully-Fisher relation. We find that smooth accretion   
of intergalactic cold gas dominates mergers for the mass growth of galaxies at $z < 2$. Next we study the baryonic   
accretion rate which shows different behaviours depending on the galaxy mass. The  
bias is also computed at different radii and epochs. Then we present   
galactocentric angular maps for the accretion integrated  
between $z=2$ and $z=0$, which reveal that accretion is highly anisotropic.  
Average accretion rates plotted against galactocentric latitude show a variety  
of behaviours. In general, accretion in the galactic plane is favored,   
even more for baryonic matter than for dark matter. Our results form a basis for   
prescribing realistic  
accretion in simulations of isolated galaxies.  
}  
  
\maketitle  
  
\section{Introduction}  
  
The most successful galaxy formation and evolution scenario is the hierarchical  
theory, which has been developped through semi-analytical models, based on the  
Press \& Schechter (\cite{Press}) formalism and its extensions   
(Bond {\sl et al.} 1991, Lacey \& Cole 1994, Kauffmann {\sl et al.} 1993, Somerville \& Primack 1999).  
In this theory, small irregular and disk galaxies   
form first, then merge hierarchically to produce larger and larger galaxies, ending  
up in giant ellipticals. The mass is dominated by collisionless dark matter,  
whose halos continue to hierarchically merge and form groups and clusters,  
In this model, the present morphology of a  
galaxy is intimately linked to the history of its interaction with the  
extragalactic environment through mergers. However other types of interactions, 
such as smooth accretion (the only contribution in the monolithic collapse model)   
and stripping, are equally important.  
  
This is true in particular to explain galaxy disks, where smooth accretion  
must prevail (e.g. Abadi et al 2003). If the hierarchical clustering picture is  
simple and successful for collisionless dark matter, the physics of baryons  
is much more complex, including star formation and feedback, and the morphology  
of present day galaxies are not dominated by mergers, although they can still  
accrete tidal debris from a previous event.  
  
This connection between morphology and environmental history can be studied  
both by observations and numerical simulations. Recently  studies
have been carried out of the fraction  
of each morphological type as a function of redshift in observed galaxies (see,  
e.g. Lilly et al, 1998, Van den Bergh {\sl et al.} \cite{Vandenbergh}). However,  
many possible observational biases make the interpretation difficult.  
In a more targeted work, Zabludoff \& Mulchaey  
 (\cite{Zabludoff}) find from observations a high proportion of disk galaxies  
 in poor clusters and infer that the merger rate is lower   
in poor clusters where galaxies represent a smaller fraction of the total mass.  
It is consistent with the picture where dynamical friction is a critical  
factor for the determination of the merger rate. This is an example of the  
crucial role of the cosmological environment on the morphological type of galaxies.  
  
Morphological features such as bars and polar rings are strongly  
dependent on the environment. Important constraints can be derived  
for example from the issue of bar formation and destruction.  
Several authors (Van den Bergh {\sl et al.} \cite{Vandenbergh}, and refs. therein)   
observed a deficiency of barred galaxies at high redshift. In a recent work,  
 however, Sheth {\sl et al.} (\cite{Sheth}) find that this deficiency could  
 be a bias due to insufficient resolution, and Jogee et al (2004) found  
evidence for a constant bar frequency with time. In this case it is necessary to  
find a process that allows bar reformation since bars are known to  
dissolve in a few Gyr (Hasan, \cite{Hasan}). Bournaud \& Combes (\cite{Bournaud02}) find   
from N-body simulations that a sufficient rate of gas accretion from  
the intergalactic medium allows bars to reform.  
  
In these two examples the competition between smooth accretion of intergalactic  
gas onto galaxies and mergers of galaxies plays a key role. The first favors  
(barred) spiral galaxy formation, the second favors the evolution toward elliptical  
galaxies. Moreover both are necessary to reproduce some morphological features.  
Kobayashi (\cite{Kobayashi}) shows from numerical simulation that both mergers  
and accretion are necessary to reproduce observed radial metallicity gradients  
in galaxies.  
  
Then galaxy mass assembly raises the following question: on average, how much of a galaxy's
mass is gained by smooth accretion, and how much by mergers? Murali {\sl et al.}
 (\cite{Murali}) give the following answer from numerical simulations: at   
$z=2$ accretion dominates mergers by a factor of 4, this value drops  
to 2 at $z=0$. This is an interesting result since accretion tends to be  
overlooked in the face of the successes of the hierarchical theory which   
emphasizes merging. Our aim in this work is to describe in detail the  
properties of accretion of both baryonic and dark matter onto galaxies in a   
cosmological framework. Aubert {\sl et al.} (\cite{Aubert}) study the geometry  
of dark matter accretion on a very large set (up to $5. 10^4$) of moderately   
resolved ($5. 10^9$ M$_{\odot}$) massive halos. Our approach is complementary;  
we look at the properties and geometry of accretion of both baryonic and dark  
matter in a small set of well resolved individual galaxies  
($6.4 10^6$ M$_{\odot}$ of baryonic mass resolution).  To achieve this   
goal we have developed a new numerical technique.  
  
To study detailed properties such as the anisotropy of accretion on a galaxy  
in a cosmological framework, we need a simulation which spans a scale range  
from 1 kpc to 10-100 Mpc. This has become possible in recent years with N-body  
treecodes or AMR codes on parallel supercomputers. In the case of N-body  
simulations the usual method is the following: a first cosmological simulation  
is performed where poorly resolved galaxies are identified, then a second  
simulation zooms in on the interesting regions (Navarro \& White \cite{Navarro94}).  
In this method as well as in AMR codes, the key idea is adaptative   
refinement of the simulation to spend CPU time where it is needed: in   
interesting regions. We present in this work a new numerical method which   
fits in the same family of adaptative refinement methods. Using a treecode, we   
perfom, for a given model, a series of $N$-body simulations (4 in practice)   
zooming recursively onto the   
studied object. The simulation box size decreases at each step while the  
resolution increases. Matter inflow and outflow and the tidal field from matter surrounding  
the current box are read in from data recorded at the previous level. Using this method  
with 4 zoom levels, we can reach the necessary dynamical range on a single  
processor workstation. Our {\sl multi-zoom} method is implemented  
for parallel computers using OpenMP.  
  
In section 2 we summarize briefly the main aspects of our tree-SPH   
implementation and of our physical model for the baryonic matter, and we  
describe in details our {\sl multi-zoom} method. In section 3 we describe a  
validation test: the collapse of a pancake. In section 4 the method is applied  
to study accretion on highly resolved galaxies. Section 5 is devoted  
to discussion and conclusions.  
  
\section{Numerical methods}  
  
Our code is designed to study galaxy formation within a cosmological   
framework with a detailed modeling of the baryonic matter physics.  
A usual Tree-SPH implementation is boosted by a new multi-zoom method  
allowing a dynamical range from 1 kpc to 10 Mpc and more, at relatively low CPU-cost.   
We model baryonic matter as a  multiphase medium; we   
 take into account two gas phases (cold and warm) and  
a stellar component, and follow energy and matter exchanges between these  
phases. Dark matter is also included with purely gravitational dynamics.  
  
\subsection{ Tree-SPH implementation - brief summary}  
The core dynamical code is a tree-SPH similar to the Hernquist \& Katz   
(\cite{Hernquist89}) inplementation.  
A detailed description of the code and a validation test (collapse  
of an initially static isothermal sphere of gas Evrard \cite{Evrard},   
Springel {\sl et al.} \cite{Springel}) can be found in Semelin \&   
Combes (\cite{Semelin02}). We use an opening criterion $\theta=0.8$ and  
quadrupole expansion for the computation of the gravitational forces with the   
tree algorithm. We keep a constant Plummer softening and a constant time  
step at a given zoom level. But we decrease both the softening length and the time  
step as we zoom in (see section 4.1). Periodical boundary conditions are   
available using Ewald method (Ewald \cite{Ewald}).  
  
In the SPH forces computation we use a spherically symmetric spline  
kernel (Monaghan \& Lattanzio \cite{Monaghan85}), arithmetic average for the  
smoothing length $h_{i,j}$, and the simple viscosity scheme described by Monaghan  
(\cite{Monaghan92}) with $\alpha=1.$ and $\beta=2.$. We use 25 SPH neighbors.  
  
\subsection{ Multi-zoom technique}  
 
\subsubsection{Algorithm}  


\begin{figure}[t]
\resizebox{\hsize}{!}{\includegraphics{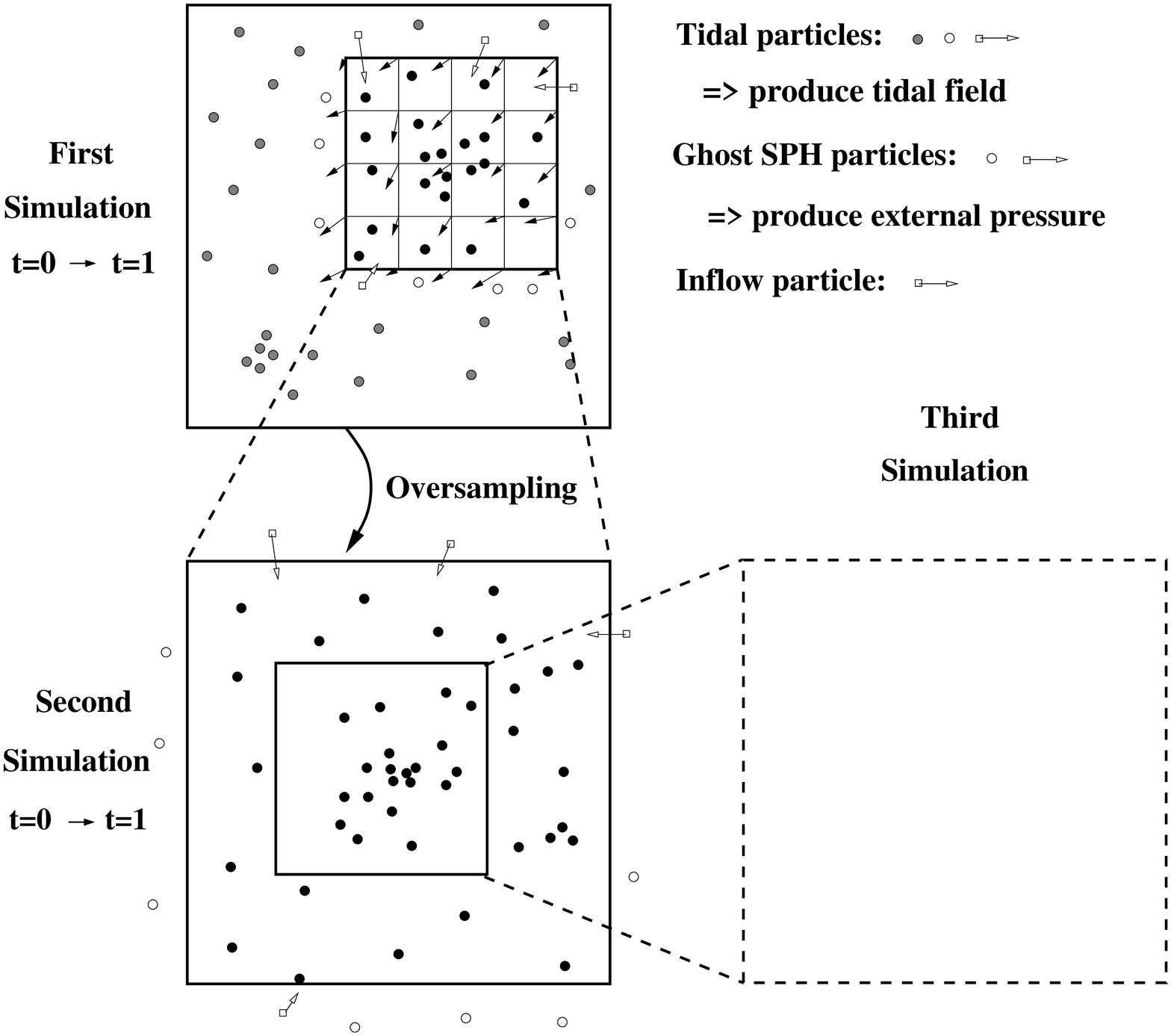}}
\caption{The multi-zoom method. The simulations are
run one after the other with increasing mass resolution in smaller and smaller
regions, using data recorded in the previous simulation to take into account
the action of matter surrounding the simulation box.
}
\label{zoom_fig}
\end{figure}

It is necessary to study galaxy formation within large scale cosmological simulations    
to take into account self-consistently such major processes as gas smooth accretion   
onto galaxies and  
mergers with other galaxies. The difficulty is to cover a dynamical range from  
less than 1 kpc, necessary for disk dynamics, to a few tens of Mpc where the   
galaxy  
distribution tends to homogeneity. This requires high mass resolution which  
comes at a high CPU cost. Specific numerical techniques have been   
developed to reduce this cost. Navarro \& White (\cite{Navarro94}) give a first  
zooming procedure which has been widely reused. A first   
cosmological simulation is run at moderate mass resolution. A galactic halo is   
identified at the present time and particles belonging to the halo are traced back   
to their initial positions, defining a zooming region. Initial conditions are   
then oversampled in this region, and undersampled outside of the region to  
retain tidal interactions from surrounding matter at low CPU cost. A second   
simulation can then be made at higher mass resolution. One limitation of this   
method is that the mass resolution difference between the two simulations cannot  
be too large. Indeed if it was,  and if we used an oversampled region tightly defined   
around the region of interest, the interaction with very massive particles outside the  
zoom region would create artificial perturbations. One way to alleviate the problem is   
to use gradually increasing mass resolution toward the center of the  
galaxy. But in practice, this requires several zooming simulations: indeed the central,  
very high resolution region boundaries cannot be correctly defined from the coarse  
cosmological simulation.  
Following this idea we have developed a new numerical method.  
  
We run a first simulation in a cosmological cubic volume and {\sl record}  
the tidal field in a sub-region at each time step. We sample the tidal  
field on a grid, at a resolution consistent with the mass resolution  
of the simulation. The tidal field is computed at each point of the grid by   
considering this point as a virtual particule and using the same tree   
algorithm as for the usual gravitational force computation. Of course, only   
particles outside the sub-region are included when we build the tree and thus   
contribute to the tidal field.  
We also record all the properties of the matter inflow  
(position, velocity, internal energy, etc...), and the positions of SPH   
{\sl ghost-particles} just outside the sub-region boundaries, which interact with  
particles inside the sub-region. We have used cubic subregions, but any shape   
can be used.  
  
\setlength{\textfloatsep}{0.4cm} 
\begin{figure*}[p]  
\centering  
\resizebox{15.5cm}{!}{\includegraphics{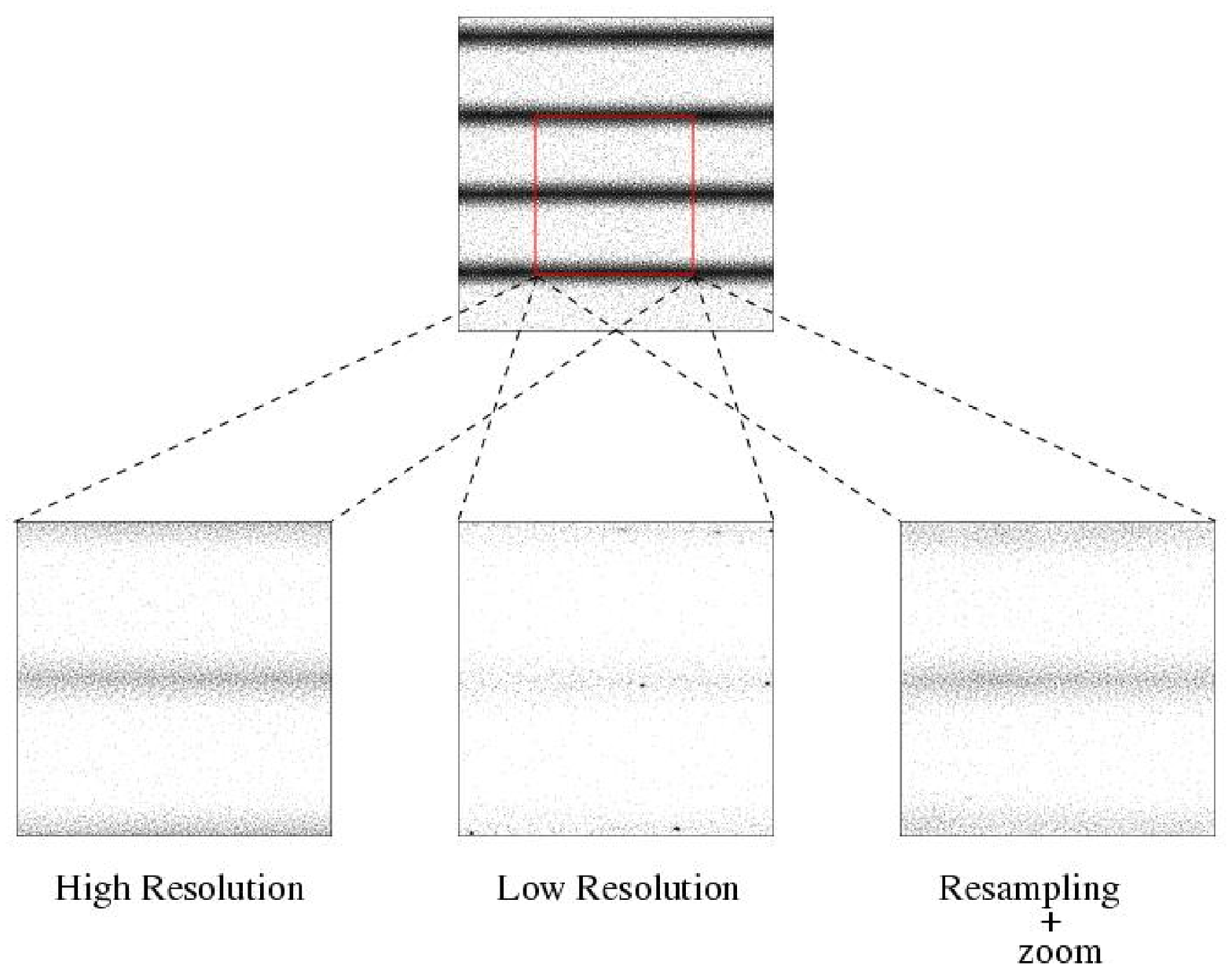}}  
\resizebox{13.5cm}{!}{\includegraphics{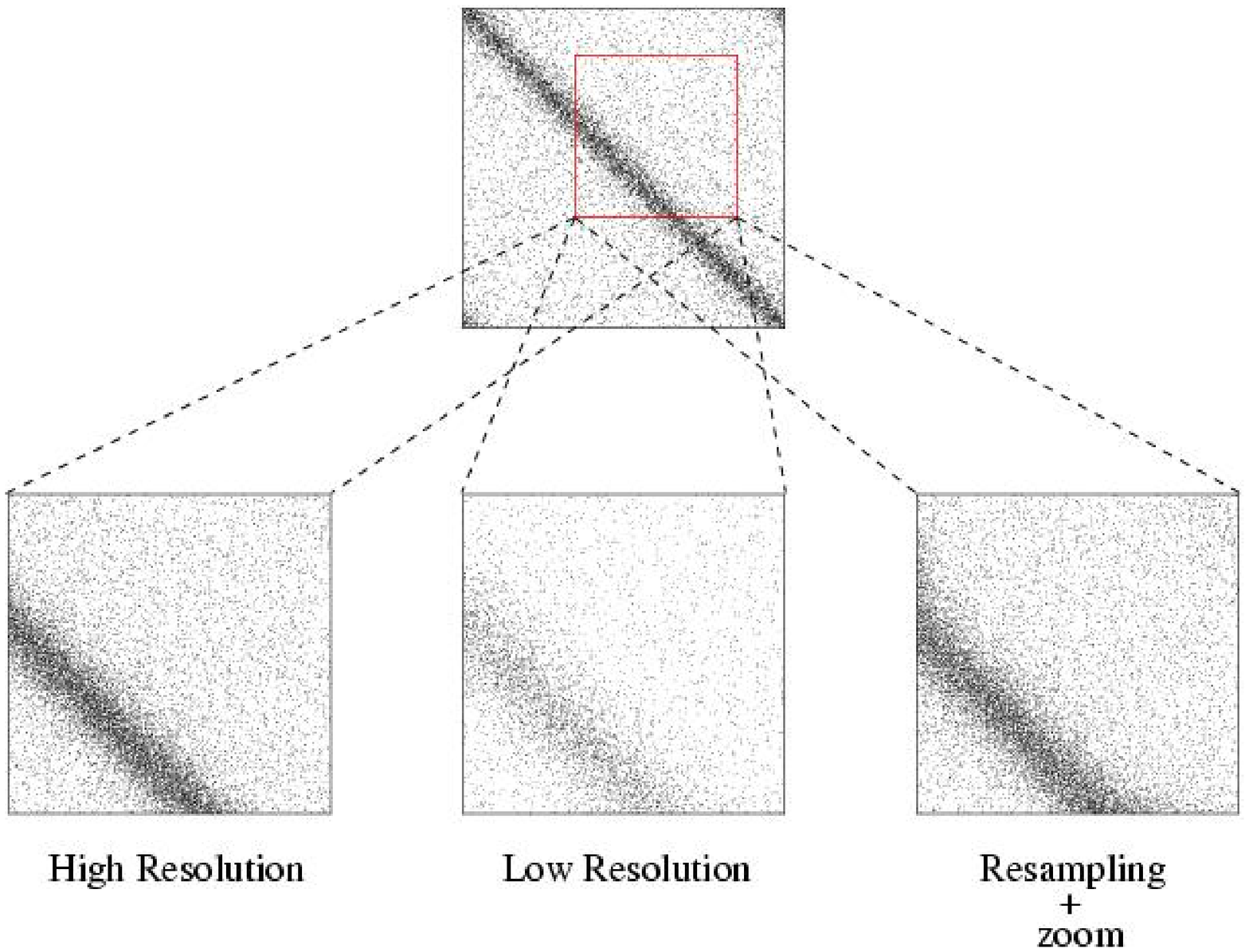}}  
\caption{Pancake formation. Particles are plotted at the instant when the  
shock steepening is maximal (see Fig. \ref{velprof_fig}). Cases of pancakes  
either perpendicular to the boundary (upper panel) or tilted (lower panel) are   
presented.  In each case, the top box shows  
the full simulation box in the high resolution simulation. The three bottom  
boxes show the zoom-in subregion in the high and low resolution simulations,  
and in the zoom simulation.  
\label{test_fig}}  
\end{figure*}  
  
Then, we run a new simulation limited to the  
sub-region which we have oversampled. The particles outside the sub-region are  
ignored: the dynamics inside the sub-region is computed taking into  
account the recorded data (tidal field, external pressure and viscosity from  
SPH ghost-particles, and matter inflow). The action of the tidal field for  
a given particle is computed by cloud-in-cell interpolation of the recorded  
grid-values. This gives very accurate values in the center of the zooming  
region, but can produce small errors near the boundaries, when small and massive  
clumps lurk just outside the region.  
SPH ghost-particles are first split to match the mass resolution of the sub-region. The newly split particles are distributed in a volume consistent with local  
 density and density gradient. Local internal energy gradient is also taken into  
 account to assign internal energies to the new particles. This procedure   
allows for smoother boundary conditions. 
Then, the split ghost particles are simply included in the neighbour search  
for the SPH particles in the subregion, and contribute, if relevant, to the pressure and viscosity.  
Their own motion however is fixed from the previous zoom level. Finally, new  
particles are introduced according to the recorded matter inflow. We have  
chosen to keep a single mass for any particle type, so each inflowing particle  
has to be split according to the mass resolution ratio between two zoom  
levels; 8 in our case. The 8 new particles are distributed around the recorded  
position in a volume consistent with the local density, and a velocity  
dispersion is added to the recorded velocity to produce a more a less  
virialized 8-particle clump.  
 
The gravitational smoothing length is reduced by the same factor as the scale  
resolution  between the initial box and the sub-region, a factor of 2 in our case. The adaptation of the time-step is a more complex issue; it depends on the  
dynamical properties of the simulated systems. In the case of large scale  
structures or interstellar clouds (above 1 Mpc or below 100 pc), there is a  
self-similar law for the free fall time: $t_{ff} \sim \sqrt{L}$ where $L$ is  
the scale. In this case the time step should be proportionnal to the square  
root of the space resolution. However, we study galaxy formation and our scale  
resolution lies in between those domains so this relation does not apply.  
Lacking a simple model for the dependence on scale of the free fall time in this  
scale range, we choose a conservative value for the time step in the  
{\sl smallest} sub-region, and  we increase it linearly with the scale resolution in larger region(s).  
  
Both tidal field and matter inflow are recorded at each time step of the   
initial box. Consequently they are updated only every 2 time steps   
(for a ratio of 2 in the time steps) in the sub-region simulation.   
At this stage, we do not make  
any time interpolation to update these data each time step. This would be  
a possible improvement in the code for the future.

We apply this procedure {\sl recursively}. In practice we have used up to 4   
zoom levels reducing the box size  
by a factor of 2 at each level, and oversampling by a factor of 8. Using  
$32^3$ simulations we obtain a mass resolution equivalent to a $256^3$  
simulation, although only over a small (but interesting) fraction of the initial volume. However,  
since we usually zoom in where galaxies are created, the inflow of matter  
can bring the number of particles up to $100^3$ at the last zoom level in a  
large cluster. Fig. \ref{zoom_fig} presents a schematic view of the method.  
  
\subsubsection{Performance and reliability}  
  
The code is a usual Tree-SPH implementation, it shows standard performances for  
such codes. The overhead connected to computing, recording and re-using the  
tidal fields and matter inflow is limited: no more than 25 \% of the total  
CPU time. The code is parallelized using OpenMP, is shows almost perfect scaling  
up to 16 processors for weakly clustered initial conditions. For highly   
clustered conditions the speed-up factor drops to 5 for 8 processors.   
This will be  
improved upon in future simulations. The comoslogical simulations in this  
work were performed on 8 processors using an IBM SP4 at IDRIS (CNRS computing center).  
  
The first obvious reliability question  is the use of a recorded tidal field.  
The oversampling produces (small) modifications of the matter   
distribution in the zooming box and we do not take into account the feedback  
of these modifications on the dynamics of the matter outside the zooming box.  
This is the main difference to Navarro \& White's method. We expect  
this feedback to be especially small in cosmological simulations in the non-linear   
regime, as long as we  
zoom in on density peaks. Indeed, in this regime, the internal dynamics of   
different clusters are largely decoupled. However, to check the robustness of   
the method, we have run a simulation in a difficult test-case where structures   
extend beyond the  
zooming box, see section 3. We have also checked that in a similar   
{\sl static} situation,  
the reconstruction of  tidal forces by interpolation produces errors smaller  
than the average error on the gravitational force due to the multipole  
expansion in the tree evaluation.  
Consequently we are confident in the reliability of our multi-zoom technique.

However, it does have some limitations. Obviously, some information is missing   
when a density structure crosses a boundary from low resolution to high  
resolution regions: fine details of the structure, absent in the low  
resolution region, cannot be recreated. Only subsequent evolution will 
produce such details. This limitation also exists in Adaptative  
Mesh Refinement codes. Navarro \& White's method is not affected by this 
limitation in the same way since the boundary moves in with the matter flow and there is no  
particle splitting. One consequence  
is that mergers will be best described if the two components have both  
formed in the last-level zooming box. In practice it puts a limit on the  
number of useful zoom-levels: about 4 in our case, for a 20 Mpc initial box.  
A larger initial box will allow more levels.  
  
\subsection{Physical model}  
  
The multi-phase physical model was detailed in Semelin \& Combes   
(\cite{Semelin02}). The basic idea is that the interstellar medium (ISM) and   
inter-galactic medium (IGM) present a multiphase   
structure at scales below  the resolution of numerical simulations. Cold dense  
gas clouds move almost as balistic particles in a warm diffuse gas medium which  
behaves as a fluid. We attempt to model this sub-grid physics with   
macroscopic recipes, as is treated star formation in    
simulations on galactic and larger scales  
(e.g. Katz 1992, Mihos \& Hernquist 1994). Multiphase models have recently  
received some attention (see, {\sl e. g.} Springel \& Hernquist   
\cite{Springel2003}, Robertson {\sl et al} \cite{Robertson},   
Marri \& White \cite{Marri}).

Let us review our model briefly. We consider four phases:   
dark matter and stars,  
subject to gravity only, warm gas, subject to gravity and hydrodynamics (SPH),  
and cold gas, subject to gravity and collisional dissipation (for justifications  
see Semelin \& Combes, \cite{Semelin02}). Warm gas can cool  
radiatively and turn into cold gas at the 10000 K limit.   
Cold gas thermal evolution is not considered. Cold gas forms  
stars at a rate fixed by a Schmidt law (index 1.5). Thermal and kinetic  
feedback from supernovae on surrounding cold and warm  gas particles is   
included: the affected  
cold gas particles are transformed back into warm gas particles. The surrounding medium  
is heated and receives a radial impulsion from the supernova. A non-instantaneous  
stellar mass-loss scheme is used (Jungwiert {\sl et al.}, \cite{Jungwiert})  
to return metal-enriched matter from the stars to the warm gas.

\section{ Pancake-formation test}  
  
\begin{figure}[t]  
\centering  
\resizebox{\hsize}{!}{\includegraphics{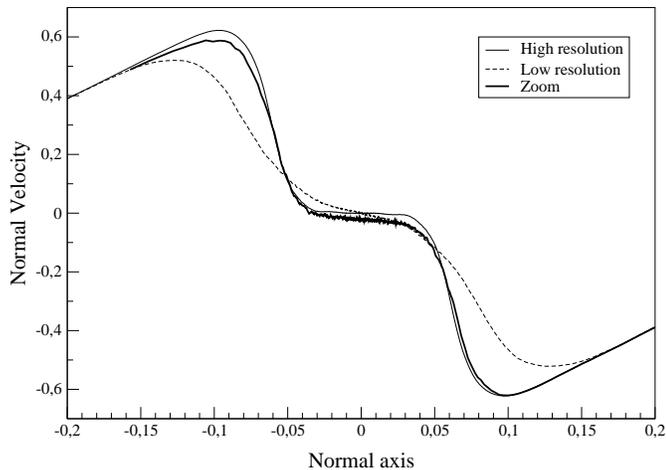}}  
\label{velprof_fig}  
\vskip -0.3cm
\caption{Velocity profile for the tilted pancake collapse test, at t=0.76,  
close to  
the instant of maximal steepness of the two inward traveling shocks. The shock  
is steeper in the high resolution than in the low resolution simulation due  
to the SPH weak shock capturing abilities. The zoom simulation reproduces the  
high resolution simulation velocity profile with very good accuracy for the right hand shock where matter has crossed no boundary. The left hand shock in the zoom simulation occurs in gas which has crossed the boundary and is affected by the low resolution dynamics.}  
\end{figure}  

To test the efficiency of our zooming method, we perform two demanding  
tests involving a strong gravitational tidal field and hydrodynamical shocks:  
the collapse of a pancake structure (Zeldovich \cite{Zeldovich}). In these  
tests initial perturbations develop and collapse in a sheet-like structure.  
  
We start with a homogeneous distribution of motionless particles in a 3D cube  
of size 1 (total mass is 1. and G=1.).  
To avoid grid effects we use the following procedure. Each particle is  
positioned at random and the system is relaxed using a repulsive gravitational force. This results in a  
glass-like homogeneous distribution. Two sets of  initial fluctuations are then  
produced. The first is obtained by applying the following displacement:  
$\delta x=-0.025 \sin(6\pi x)$ . This produces 3 pancakes perpendicular to the box boundary. The second is obtained by applying a displacement  
$-0.025 \sin(4\pi(x+y))$ to both $x$ and $y$ axis. This produces a diagonal pancake.  
Periodic boundary conditions are used.  
Cosmological expansion is ignored for this test. The initial internal energy  
is $0.01$ in the first case and $0.03$ in the second case (to counter a higher mass in the single pancake), and radiative cooling  
is turned off (adiabatic evolution). In the first case, high resolution is  
$50^3$ and low resolution $25^3$ gas-particles. In the second case, high  
resolution is $64^3$ and low resolution $32^3$ gas-particles.  
Then we zoom in on the central pancake in the low resolution simulation,  
in a cubic subregion of size 0.5. The zoom region is centered on the pancake in the first case, and off center in the second.  
In both cases, we resample the initial conditions to have the same mass resolution than in the high resolution simulation and use the recorded tidal field and  
inflow. The result of the simulations can then be compared in the central subregion.

Fig. \ref{test_fig} displays the particle distribution for the three  
simulations at t=0.85 in the fist case and t=0.76 in the second case.  
In the perpendicular (first) case, the pancake structure in the zoom simulation shows no  
noticeable defect. In the tilted case, the structure is globaly reproduced,  
but two small defects appear: two overdense tails extend out of the pancake  
along the boundaries on the side facing the center of the box. Additionnal simulations have shown that these tails are due to pressure forces: they disappear  
for dust or very low internal energy. Indeed matter in these regions feel the  
pressure of the pancake outside the sub-region, which, computed at a lower  
resolution, has a shallower profile. This is an unavoidable effect of the resolution transition. This defect is greatly enhanced by the fact that the structure  
remains motionless on the boundary: the effect builds up over time  
on the same particles, preventing then from flowing out. This situation in very unlikely in cosmoslogical simulation where we zoom on collapsing regions where  
structures flow through the boundary. In the test, no defect appears in the  
structure of the pancake on the side where inflowing gas accretes.

In these tests, the most sensitive diagnosis is the  
velocity profile, which exhibits hydrodynamical shocks. Fig. \ref{velprof_fig}  
shows the velocity profile at time t=0.76 in the case of the tilted pancake, the most difficult one. This instant is chosen to be close to the  
instant of maximal steepness in the shock fronts. Let us first notice that  
the two shock fronts are shallower in the low resolution simulation than in  
the high resolution simulation: its limited shock capturing ability is a  
well-known drawback of the SPH method.  
The zoom simulation gives a velocity profile which is very close to the high  
resolution simulation for the right hand shock where accreting gas comes from  
inside the sub-region. The left hand shock is a little bit less well reproduced. Indeed it accretes gas from outside the sub-region, whose earlier dynamics was  
computed at lower resolution. The resulting zoom shock is  a little bit shallower than the full high resolution shock.  
From this diagnosis, we would say that the zooming method reproduces velocity profiles quite well even in region of inflowing gas. What about velocity  
dispersion?  

\begin{figure}[t]
\centering
\resizebox{\hsize}{!}{\includegraphics{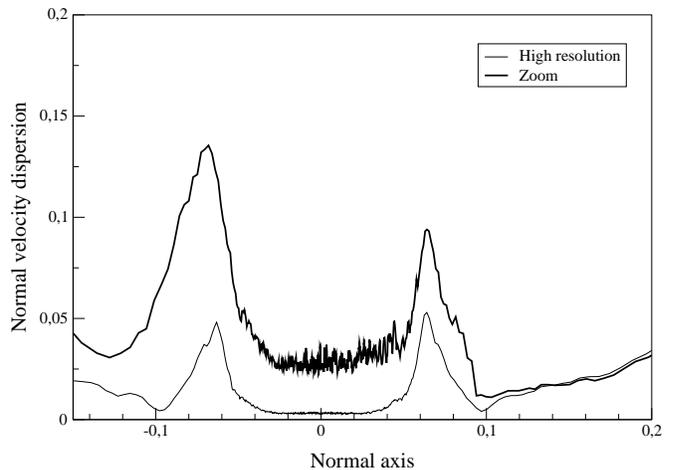}}
\label{velsig_fig}
\vskip -0.4cm
\caption{Velocity dispersion profiles in the tilted pancake collapse test.
The zoom simulation shows a higher velocity dispersion than the
full high resolution simulation in the pancake region.}
\end{figure}

In Fig. 4 we can see the velocity dispersion profiles for the high resolution simulation and zoom simulation. The dispersion  
is higher in the zoom simulation than in the high resolution simulation by a  
factor of 2 to 3 in the shock regions, and 5 or more in the center of the pancake. In the center of the sub-region 
(i.e. coordinate $\sim 0.14$ on the axis normal to the pancake in Figs 3 and 4),
far from the boundary and not directly within the hydrodynamical shocks, 
the two velocity dispersions are almost identical. 
We believe that both the resampling of inflowing particles and the noisy (unrelaxed) distribution of ghost SPH particles are responsible for the increase in the velocity dispersion.  
The density structure itself however is not visibly affected by the velocity dispersion difference, which is important for galaxy formation.

\section{Galaxy formation: properties of the accretion}  
  
We present now an application of our multi-zoom method to galaxy formation,   
focusing on the accretion of intergalactic matter onto galaxies. Our   
goal is to describe the geometrical properties of baryonic and dark matter   
accretion for well resolved galaxies. We achieve the  
necessary scale range  by using 4 zoom levels. To obtain the same range on a  
single level, we would need $256^3$ particles, and a supercomputer to run the  
simulation. With the multizoom method each 4 zoom level series of simulations   
(one  
per investigated region) required about 200 to 300 hours of single processor CPU   
times. A single zoom level in a particular region holds from $32^3$ to $100^3$   
particles depending on the matter inflow during the time of the simulation.  
  
\subsection{Cosmological model and initial conditions}  
  
We use a flat $\Lambda$-CDM cosmological model with $\Omega_{\Lambda} = 0.7$,  
$\Omega_0=0.3$. The Hubble's constant at present time   
is $H_0= 70$ km s$^{-1}$ Mpc$^{-1}$, and the baryon density   
$\Omega_{\hbox{bar}}=0.05$.   

The power spectrum of primodial density   
fluctuations is $P(k)\propto k^{-2}$, normalized to $\sigma_8=0.8$.  
The power spectrum is predicted  
in a $\Lambda$CDM model to vary with scale $k$, from  
n=1 at large scale, until n=-3 at very small scale.  
From linear theory, and starting from a scale-invariant  
spectrum, as predicted by standard theories of inflation, it is  
possible to compute P(k), which is in the above hypothesis,  
very close to P(k) $\propto k^{-2.5}$ for $0.1 < k (\hbox{Mpc}^{-1}) < 10$  
and close to  $\propto k^{-1}$ for $0.01 < k (\hbox{Mpc}^{-1}) < 0.1$  
(Klypin {\sl et al.} \cite{Klypin}).  
We adopted a constant slope of n=-2, fitted  
to the range of scales that we simulate.  
This approximation is not crucial in the computations  
done here, since we are in the highly non-linear regime,  
where the evolution modifies significantly the initial slope;  
in particular, non-linear effects will increase power on  
scall-scale, especially strongly for n $< -1$ (Peacock \& Dodds \cite{Peacock}).

We apply the usual procedure to produce the initial conditions at the first   
zoom level. We put particles on a grid and we compute their displacement and   
velocity according to Zel'dovich approximation to produce the desired   
fluctuation power spectrum, at the initial redshift of the  
simulation, $z=45$ in our case. A simple way to produce the initial conditions   
for the successive zoom levels would be to compute the initial displacement   
field in the whole box at the resolution of the last zoom level, and extract   
initial conditions for all zoom levels from this field.  
Although this would be possible in our application with 4 zoom levels   
($256^3$ grids to manipulate), it would become prohibitive in term of CPU   
with a few more zoom levels, or with a larger number of particles in the first  
level. Consequently we compute initial conditions   
recursively from first to last zoom level: at each level we read in   
displacements due to large scales stored at the previous zoom level, we add   
the displacement from the new small scales introduced at this zoom level,   
and finally we store the displacement for the next zoom level. Of course, we   
ensure that the added fluctuations at each zoom level follow the desired power   
spectrum. The advantage of this method is that the cost is linear in the   
number of zoom levels.

\begin{figure}[!t]
\centering
\resizebox{\hsize}{!}{\includegraphics{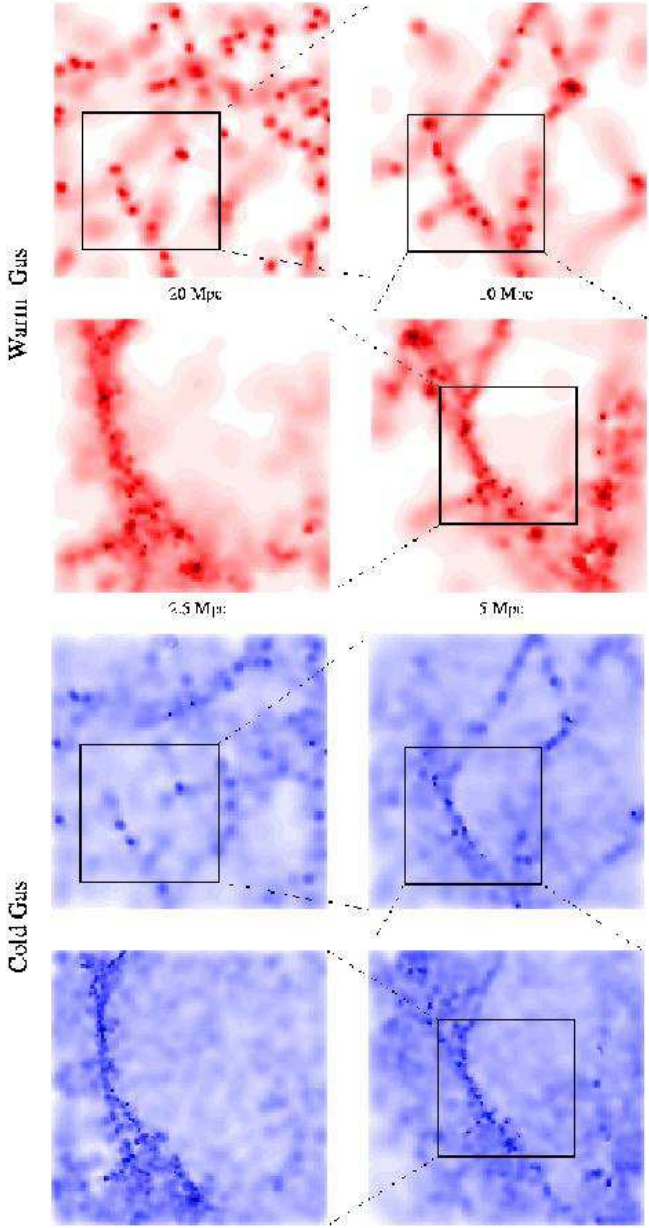}}
\caption{Warm and cold gas phases density fields over 4 zoom levels (20, 10,
5 and 2.5 kpc box sizes) at $z \sim 1.57$. The focus is on the region where
galaxies number 4 and 5 are found (see Table 1).
The density fields are computed from the gas particle positions using
an SPH-like adaptative smoothing. White cores represent high density regions
and dark areas are voids. The different dynamics of warm and cold gas (SPH vs
sticky particles) produce different density structures: rather diffuse for the
warm gas, more clumpy for the cold gas.
\label{hg_cg}}
\end{figure}
  
At the first zoom level we use $32^3$ particles, half baryons,   
half dark matter, distributed in a $20 h^{-1}$ Mpc box. This gives a baryonic   
mass resolution of $ \sim 3.3 \,\, 10^9$ M$_\odot$.  
Then, at each zoom level, we reduce the box size by a factor of 2 and   
increase the mass resolution by a factor of 8. The baryonic mass resolution at   
the fourth zoom level is $ \sim 6.4 \,\, 10^6$ M$_\odot$. Since we focus  
the zooms on the local overdensities where the galaxies form, we end up at   
the $4^{th}$ zoom level with simulations involving $10^5$ to $10^6$ particles,   
due to matter inflow. For zoom level 1 to 4, we use a gravitational softening   
of 20, 10, 5 and 2.5 $h^{-1}$ Mpc and a timestep of 10, 5, 2.5 and 1.25  
Myr.  
  
\subsection{ Gas cooling in a cosmological framework}

One of the focus of our simulations is the separation between a warm and a   
cold gas phases with different dynamics. In Semelin \& Combes (\cite{Semelin02})  
we have studied in details the exchanges between the two phases within a   
galactic disk. This study is also relevant at cosmological scales, but new   
behaviours appear. First, at high redshift, adiabatic cooling due to   
cosmological expansion dominates the cooling of warm gas. This mechanism is  
so effective that, if not counterbalanced by first star formation, all the  
warm gas turns into cold gas early on. Since the mass resolution is not sufficient to  
produce early star formation leading to the reionization of the universe,   
the reionization is introduced artificially using a constant heating of   
cold gas particules (those below $10$ K) between $z=15$ to $z=6$. Cold gas   
will then replenish the warm gas phase, and a sizeable fraction of warm gas  
survives the fast-expansion era.  

To summarize, warm gas   
is turned into cold gas both where collapse occurs in high density regions   
(through radiative cooling) and in low density, dynamically quiet   
regions (though adiabatic expansion). Warm gas survives mostly in dynamically   
active (turbulent) regions around density peaks, where it is heated first by  
reionization, and later by   
hydrodynamical shocks. It forms a diffuse halo. The different   
morphologies of warm and cold gas phases can be seen on fig. \ref{hg_cg}.   
The warm gas phase is also replenished at smaller redshift  
by star formation feedback. However, without a proper treatment of radiative   
transfer including the propagation of ionisation fronts, this feedback energy   
is closely confined around star-forming galaxies and is not transported into   
the inter-galactic medium.

\subsection{Galaxy identification and basic properties}  
 
\begin{figure}[t]
\centering
\resizebox{\hsize}{!}{\includegraphics{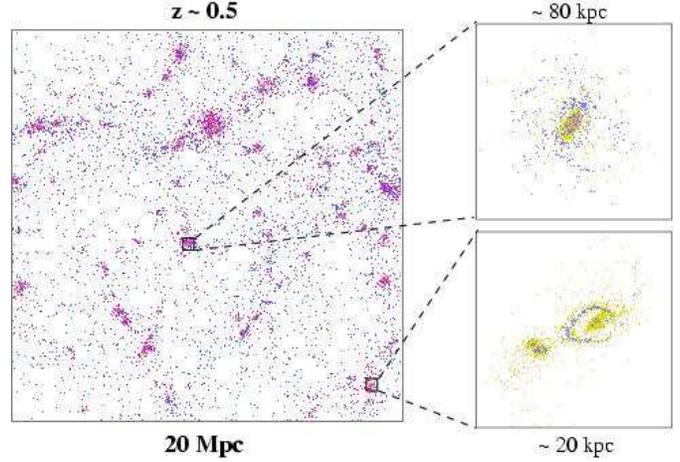}}
\caption{Example of field galaxies selected for our study (number 1 on top and n
umber
 8 at bottom) at $z=0.5$.
Magenta particles are dark matter particles, blue particles are cold gas particl
es
and yellow particles are stars (electronic version only). Dark matter is not plo
tted
in the two zooms. Galaxy 1 (face-on) exhibits an
extended cold gas disk, and a stellar disk growing from inside out.
There is little bulge to speak of: the extent of gravitational softening prevent
s its formation.
Galaxy 8 is undergoing a merger.
\label{zoomin}}
\end{figure}

We have investigated 8 regions down to the fourth zoom level. Among them, 6   
regions provide good galaxy candidates. The region where positioned {\sl by   
hand}. Then we apply a clump finding algorithm to the 6 regions. We use a   
{\sl friend of friend}-like algorithm (e.g. Huchra \& Geller, 1982).   
We compute a SPH density for {\sl all}   
(even non-gas) particles. Then we select local density maxima (particles with   
a higher density than their SPH neighbours), and starting with the highest   
maximum, we build the clump by accreting neighbouring particles with lower  
density (friend of   
friend part) which are above a given density threshold (the average density)   
and do not already belong to another clump. Then we remove from the local   
maxima list the clumps which are below  
a given number of particle threshold (500 particles here). We iterate the   
process 5 times, re-sorting the order in which we examine the clumps by   
decreasing mass. We applied this algorithm to the six regions at $z=2$, and   
identified 10 galaxies with a mass within a factor 10 of the Milky Way mass.   
We rejected dwarf galaxies since we want $\sim 10^4$ baryonic particles per  
object.  Another creterion for rejecting galaxies was crossing, or being too   
close to the boundaries of the zooming box. Indeed we wish to minimise the  
impact of accretion of matter which underwent a change in mass resolution.   
  
The masses of the selected objects at $z=0$ are given in Table 1.  
Two examples of this selection process are shown on fig. \ref{zoomin}.  
  
\begin{table}[t]  
\centering  
\setlength{\tabcolsep}{0.1cm}  
\begin{tabular}{|c||c|c| c|c|c|}  
\hline  
Galaxy & Group& Baryonic & Gas &$\tau_{\hbox{ac}}$ &Mass growth \\   
       & &mass in $M_\odot$   & fraction & in Gyr& by mergers\\  
\hline  
\# 1 &L& $1.0 \,10^{12}$& 19.1 \%& -& $< 5\%$\\  
\# 2 &S& $6.2 \,10^{10}$& 2.2 \%&1.7& $ 21\%$\\  
\# 3 &L& $1.9 \,10^{12}$& 24.4 \%& -& $ 25.5\%$\\  
\# 4 &M& $4.0 \,10^{11}$& 12.6 \%& 4.7& $ 55\%$\\  
\# 5 &S& $3.7 \,10^{10}$& 3.6 \%& 1.3&$ 9\%$\\  
\# 6 &M& $1.9 \,10^{11}$& 16.5 \%& 21.4&$ 40\%$\\  
\# 7 &S& $9.0 \,10^{10}$& 10.0 \%& 4.4&$ 7.5\%$\\  
\# 8 &M& $3.4 \,10^{11}$& 7.3 \%& 8.9&$ 12.5\%$\\  
\# 9 &M& $3.1 \,10^{11}$& 16.0 \%& -&$ 25\%$\\  
\# 10 &L& $1.1 \,10^{12}$& 11.5 \%& -&$ 63\%$\\  
\hline  
\end{tabular}  
\vskip 0.2cm
\caption{Basic properties at $z=0$ of identified galaxies.   
They are classified in 3 groups   
according to final mass, S for small, M for medium and L for large.   
The baryonic mass is computed within a 50 $h^{-1}$ Mpc radius at present time.  
The gas fraction is the percentage of hot+cold gas to total baryonic matter  
within a 50 $h^{-1}$ Mpc radius at present time. $\tau_{\hbox{ac}}$ is the decay time of smooth accretion rate for baryonic matter (see main text). The mass growth from mergers is the fraction of baryonic mass acquired between z=2. and z=0. by   
mergers with a mass ratio greater than $\sim 1 \over 20$.  
\label{tabmass}}  
\end{table}  
\begin{figure}[t]  
\centering  
\resizebox{\hsize}{!}{\includegraphics{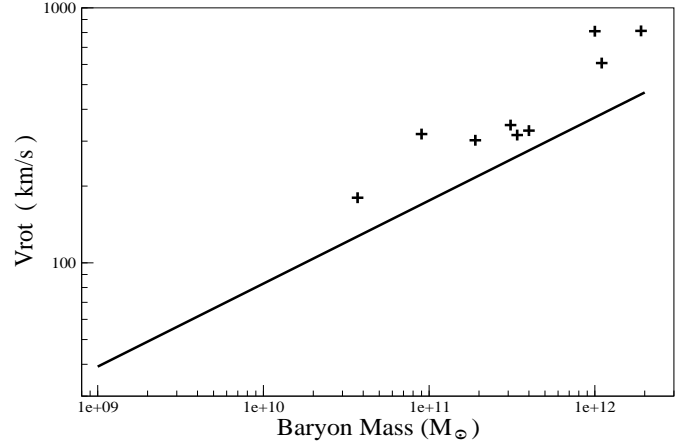}}  
\caption{Tully-Fisher relation for 9 of the 10 identified galaxies in the sample, at $z=0$.   
The maximum rotation velocity in the gaseous disk is plotted against the  
 baryonic mass of the galaxy. The full line represents a fit through observed galaxies  
(e.g. Giovannelli et al 1997).  
\label{tully}}  
\end{figure}  
Let us make a few remarks about the disk structures formed in the simulation  
which can be seen in Fig. \ref{zoomin}. First, cold gaseous disk do form, and  
have spiral arms. A stellar disk forms from inside out. Warm gas is only  
present through the local feedback of star formation. Fig. \ref{tully} shows  
the Tully-Fisher relation for 9 of the 10 galaxies (the last galaxy shows no  
definite rotation). The maximum rotation velocity is computed by a procedure  
similar to observations: an histogram of velocities along a line of sight  
is computed for gas particles, the inferred rotation velocity is half the width  
of the histogram at 20 \% of the maximum value. This procedure is repeated for  
a large number of random lines of sight, and the maximum value of the velocity  
is retained. The law fitted from observations is plotted as a full line. The  
slope is taken from Giovanelli {\sl et al.} (\cite{Giovanelli}), 0.325 on our  
graph, and the zero point is the Milky Way with baryonic mass $2. \, 10^{11}$   
M$_{\odot}$ and maximum rotation velocity $220$ km.s$^{-1}$. This plot should be  
interpreted with care since all but the 3 largest galaxies have disks with radii  
smaller than 3 times the gravitational softening.  
With this in mind we can check that, the Tully-Fisher  
relation of the numerical galaxies has a correct slope but a shifted zero  
point, which is similar to the detailed study by Navarro and Steinmetz (\cite{Navarro00}).   
The rotational velocity excess is connected to the angular momentum  
problem: disks are too small because of a large loss of angular momentum towards  
the dark matter halo.   
The problem may be alleviated and large disks may be   
obtained by modifying the equation of state for the gas, as suggested by   
Robertson {\sl et al.} (\cite{Robertson}).   
  
In any case, accretion at more  
than 50 kpc from the galactic center, which is the main focus of this paper,  
 is hardly affected by the disk size.

\subsection{ Baryonic mass accretion rates}  

\begin{figure*}[t]
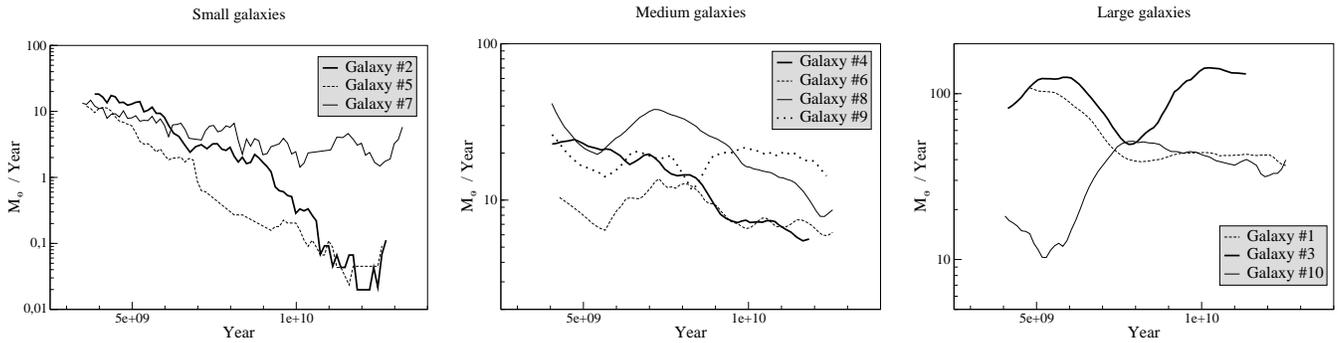

\centering
\begin{tabular}{ccc}
\resizebox{5.6cm}{!}{\includegraphics{small-gal-accr-rate.eps}}&
\resizebox{5.6cm}{!}{\includegraphics{medium-gal-accr-rate.eps}}&
\resizebox{5.6cm}{!}{\includegraphics{large-gal-accr-rate.eps}}\\
\end{tabular}
\caption{Baryonic accretion rate in $M_{\odot}$ per year for sample galaxies in
group S, M and L. Mergers have been removed from the plots. The curves have been
smoothed over 100 Myr periods.
\label{figmass}}
\end{figure*}

From $z=45$ to $z=0$, we produce 108 outputs for each investigated region.   
Using the outputs we can compute the mass accretion rates with a 125 Myr   
time resolution. For each output, the baryonic mass of the 10 galaxies is   
computed within a 50 $h^{-1}$ Mpc radius. Only particles which are also present   
in the last output contribute, to avoid the contribution from bodies on   
hyperbolic orbits.   
  
Let us first comment on the fraction of mass gained through mergers. Table \ref{tabmass} gives rough estimates (derived from sharp steps in mass growth   
histories) of the contribution of mergers between $z=2.$ and $z=0$. Only   
mergers with mass ratio larger than $\sim {1 \over 20}$ are identified in this   
estimate. Bournaud {\sl et al.} (\cite{Bournaud04}) find that mergers with a mass  
ratio smaller than $\sim {1 \over 10}$ have very little effect on the morphology  
of the large galaxy because the small galaxy is ripped appart before hiting  
the large one. Consequently mergers with mass ratio of $\sim {1 \over 20}$ and lower indeed qualify as smooth accretion.  
Although we find a wide scatter in the contribution of mergers to the mass growth, from a few percent to $63 \%$, our   
results   
agree on average with values given by Murali {\sl et al.} (\cite{Murali}). The  
bottom line is that, in general, smooth accretion dominates mergers as far as mass growth is concerned.

Fig. \ref{figmass} shows the smooth  
accretion rate histories in solar masses per year. There is one graph for each group,   
Small, Medium and Large, and one curve for each galaxy. Mergers, which  
produce spikes in the accretion rate history, have been removed leaving only  
smooth accretion.   
  
First, the average gas accretion rates computed for our sample of galaxies,   
typically in the range    
$1-100 M_{\odot} \hbox{yr}^{-1}$ for galaxies with a final mass from   
$3. \, 10^{10} M_{\odot}$ to $2. \,10^{12} M_{\odot}$ are in the same range  
as observed for star formation rates. Moreover, visual inspection of the output snapshots shows that the smoothly accreted baryonic mass is almost exclusively  
cold gas.  
This suggest that accretion replenishes the gas phase in the disk, and  
may actually regulate the star formation rate. Finally let us mention that  
our computed accretion rates are sufficient to produce the reformation of   
bars according to Bournaud \& Combes (\cite{Bournaud02}).

The accretion rate of the galaxies in the S group show a decreasing   
behaviour of the type   
$ \dot{M}_{\hbox{ac}}=\dot{M}_0 \exp(-{t \over\tau_{\hbox{ac}}})$.   
The fitted values for the typical decay time $\tau_{\hbox{ac}}$   
are given in Table 1. Galaxies in group M, except one,  show the same behaviour  
with larger values for $\tau_{\hbox{ac}}$. Galaxies in group L have  
a more or less constant accretion rate (no systematic decay).   
Exponential decay type laws for the accretion rate suggest a simple model: the  
accreted matter is taken from a finite surrounding reservoir in 
quasi-equilibrium,  at a rate proportional to the gas remaining in the
reservoir. In this picture:  
it is slow angular momentum loss which drives accretion onto the galaxy   
(which leads to the angular momentum problem, Navarro \& Steinmetz \cite{Navarro00}).   
The process could be self-regulated: the presence of more gas in the galaxy   
triggers non-axisymmetric instabilities, such as bars, which drives the gas   
inwards. D'Onghia \& Burkert (\cite{Burkert04}) find a different picture based on  
dark matter simulations where matter falls on eccentic orbits with little   
angular momentum in the first place; too little in fact to account for observed  
galactic spins even without angular momentum loss during accretion. A possible   
answer is that the dissipative dynamics of accreting baryons   
concentrates angular momentum more efficiently: the baryonic matter ending up   
in the galaxy is drawn from a larger reservoir than non-baryonic matter, thus  
has a higher initial specific angular momentum.

If star formation activity is a good tracer of the accretion rate,   
as we believe, these accretion rate histories  
are at odds with observations: real large galaxies in the center of clusters have  
little star formation activity at the present time, while dwarf galaxies keep   
forming stars at a good rate, relative to their mass. We find the opposite. We   
suggest two phenomena to explain this discrepancy. It appears that,   
in the simulations, the accretion  
rate of the smallest galaxies decays too fast: this may well be caused by the  
well-known overmerging phenomenon. The smallest galaxies are the closest to the  
mass resolution limit, where overmerging is predominant; thus, they accrete  
all the available mass early on, and, because of their low mass, are not able  
to capture any more mass later on. In other words, they build up to their  
maximal mass too fast, then the accretion stops. As for large galaxies,  
which keep accreting too much mass in our simulations, the problem may arise  
from an inadequate heating of gas at large-scale. In real clusters, we observe a   
massive virialized hot gas halo (holding as much or even more baryonic mass   
than the galaxies,  
{\sl e. g.} David {\sl et al.} \cite{David}), sustained by pressure, and   
only in the center, gas is able to cool down and infall    
onto the central galaxy (Fabian 1994). The importance of the feedback from  
the central AGN to moderate the cooling flow has now been realised  
(e.g. Brighenti \& Mathews, 2003), and this feedback is not included in our  
simulations. Moreover, in rich groups and clusters, the hot gas halo (ICM)  
is stripping the interstellar medium of galaxies, and prevents cold  
gas accretion, quenching star formation (Poggianti et al 1999).  
If such  
a halo is indeed present in the simulations, it is not massive enough.   
Taking into account radiative transfer  
and ionization fronts could alleviate the problem.  
Moreover, our spatial resolution is not enough to deal properly  
with ram pressure, which would contribute to gas stripping in   
rich environments, where massive galaxies are found.

We have shown that the average accretion rates have the expected values, but   
that accretion rate histories versus mass are affected by the limitations of the simulation.  
This, however, should not change drastically the geometrical properties of   
accretion that we study in section 4.6.

\begin{figure}
\resizebox{\hsize}{!}{\includegraphics{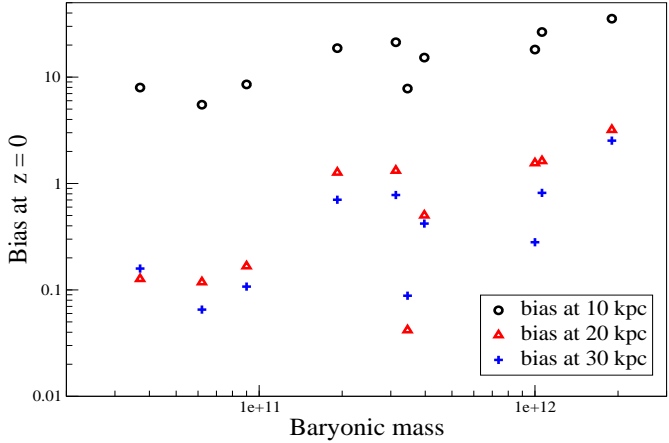}}
\caption{The local bias, baryonic to dark matter mass ratio relative to cosmic
abundance (ratio of 0.2), is plotted for each galaxy in our sample, with the mas
ses computed within a 10, 20 or 30 kpc sphere, at $z=0$.
\label{local_bias}}
\end{figure}
 
\subsection{Bias}  
  
Numerical simulations have shown that the standard  $\Lambda$-CDM model leads to  
overconcentrated dark matter halos compared with observations (Burkert \& Silk  
\cite{Burkert}, Moore {\sl et al.} \cite{Moore}). The  
angular momentum problem (baryonic matter losing momentum to dark matter,  
resulting in too small galatic disks) has also proven difficult to overcome  
(Navarro {\sl et al.} \cite{Navarro95}). The  
latter may arise from numerical issues only, but both are connected to matter  
accretion onto forming galaxies. A crucial diagnostic for the distribution of  
matter around galaxies is the local (those below $10$ K) bias: that is, the local baryonic matter to  
dark matter mass ratio relative to the average cosmic ratio of   
$\Omega_b/\Omega_{CDM}$ = 0.2. We have computed  
this diagnostic for the galaxies in our sample.

For each galaxy, we have computed the total mass of dark and baryonic matter  
within 10 kpc, 20 kpc and 30 kpc from the galatic center at $z=0$. Then we  
have derived the corresponding local biases, which are plotted in fig.  
\ref{local_bias} against the baryonic mass of the galaxy as computed in Table  
1. The first obvious effect is that bias decreases as radius increases:  
this is to be expected since baryonic matter is highly concentrated within  
dark matter halos.  
Next we can check that at a given radius, the bias increases with the galaxy  
mass. This is in line with observations which show that dwarf galaxies are  
dominated by dark matter (small biases) while intermediate and giant
galaxies are not. The  
biases at 10 kpc which are less noisy than biases at 20 and 30 kpc (due to  
accreting gas clumps without dark matter haloes) even  suggest a  
power law relation between the bias and the galactic mass. The quantitative  
value of the exponent may depend on the particular model and strength of  
gas dissipation. It requires a more thorough study.  
  
\begin{figure}{!}
\resizebox{\hsize}{!}{\includegraphics{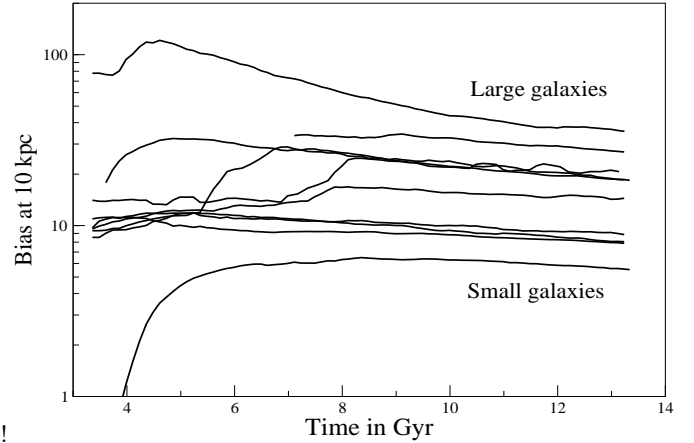}}  
\caption{The evolution of the bias at 10 kpc (see def in Fig. \ref{local_bias})  
of all the galaxies in our sample is plotted against time. Individual galaxies  
are not identified since we are looking for a common global behaviour.   
Large galaxies are at the top, small galaxies are at the bottom.  
\label{bias_evol}}  
\end{figure}

In Fig. \ref{bias_evol} we plot the evolution with time of the bias at 10 kpc  
for all the galaxy. Since we are interested in a common behaviour we did  
not identify individual galaxies on the plot, but as Fig. \ref{local_bias}  
shows, higher biases are assiciated with larger baryonic mass. The salient feature  
of Fig. \ref{bias_evol} is that, after a strong initial rise, the bias  
slowy decreases with time. Baryonic matter accretes very fast during the  
initial formation era of the galaxy mass assembly,  
 then drops and even stops in some cases  
at later times, while the dark matter halo keeps concentrating.  
The bias decay rate seems independent of the galactic mass  
at late times (between 10 and 13.7 Gyr). We computed the decay times;  
they change by a factor of more than 3 in the range of a few tens of Gyr.

\subsection{ Accretion maps}  

\begin{figure*}[t]  
\centering  
\begin{tabular}{cc}  
\resizebox{7.5cm}{!}{\includegraphics{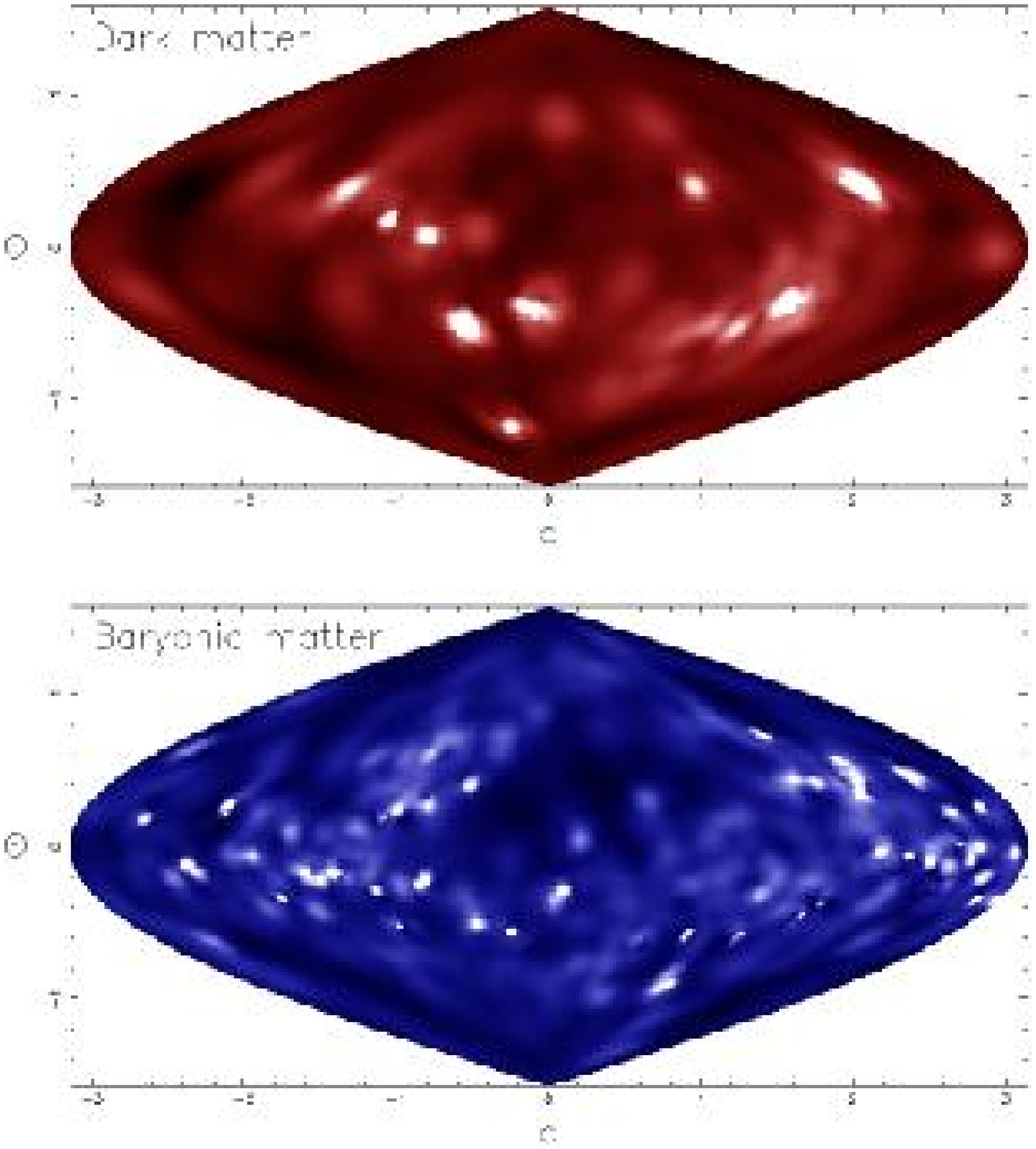}} &  
\resizebox{7.5cm}{!}{\includegraphics{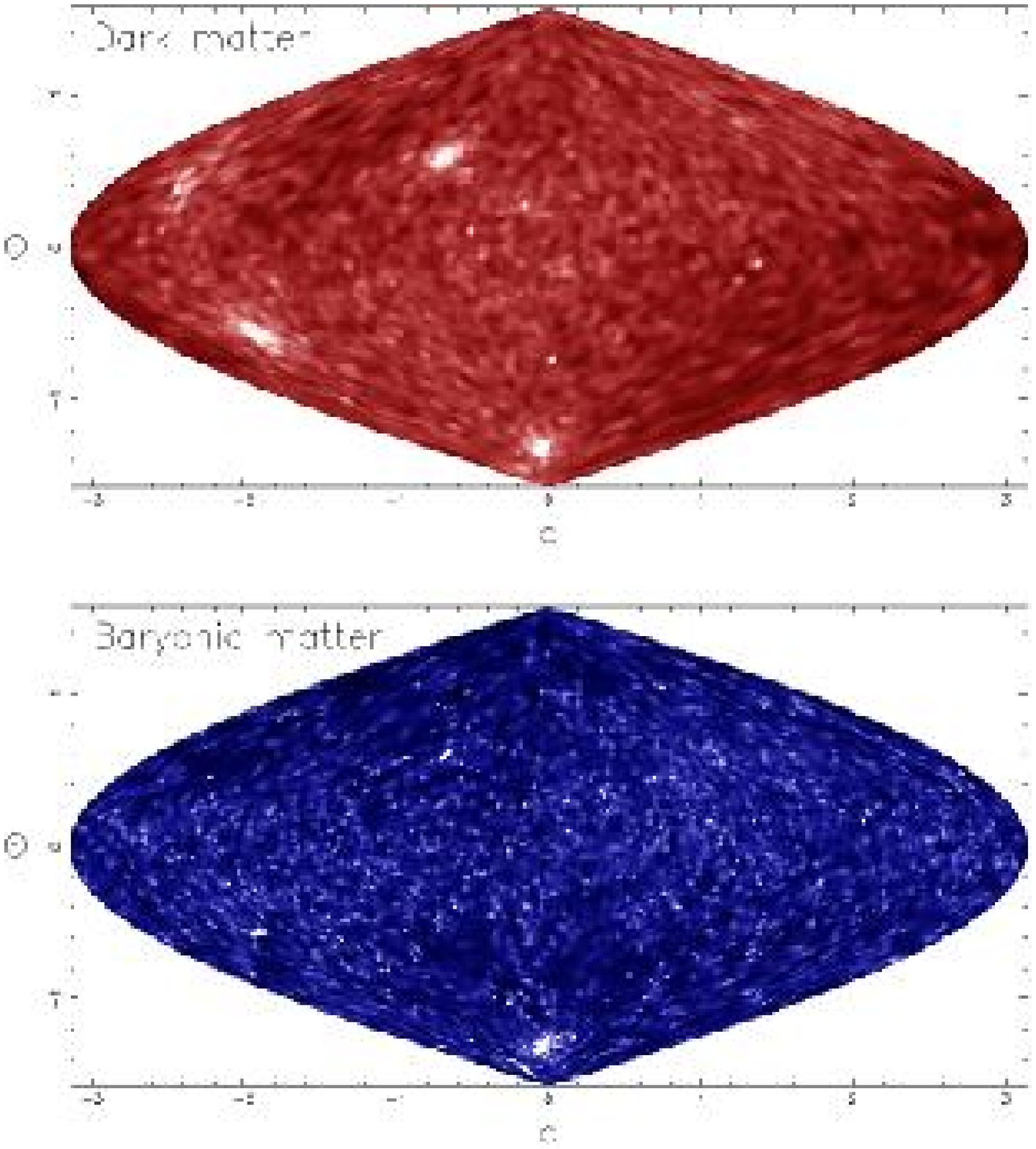}} \\  
\resizebox{7.5cm}{!}{\includegraphics{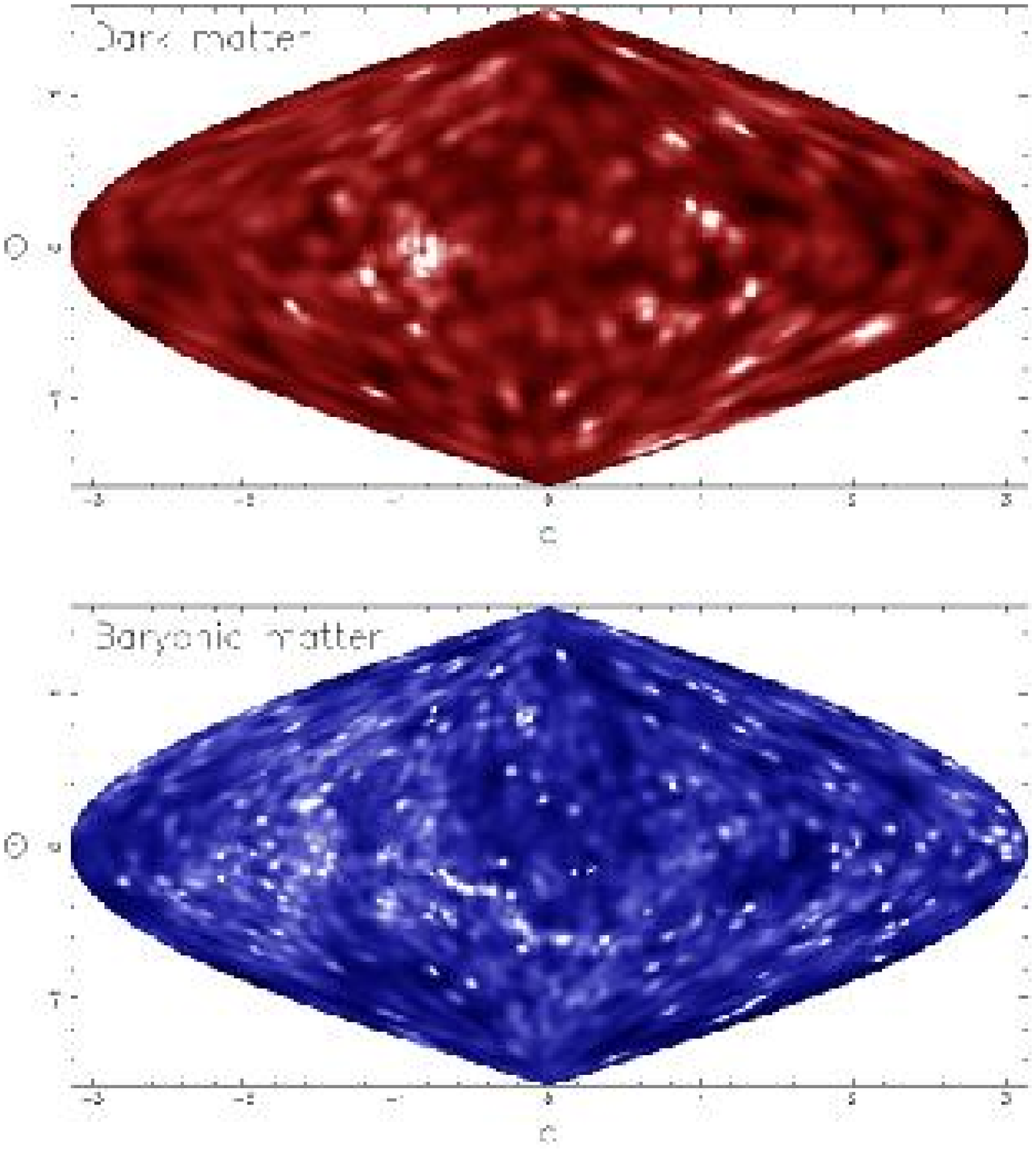}} &  
\resizebox{7.5cm}{!}{\includegraphics{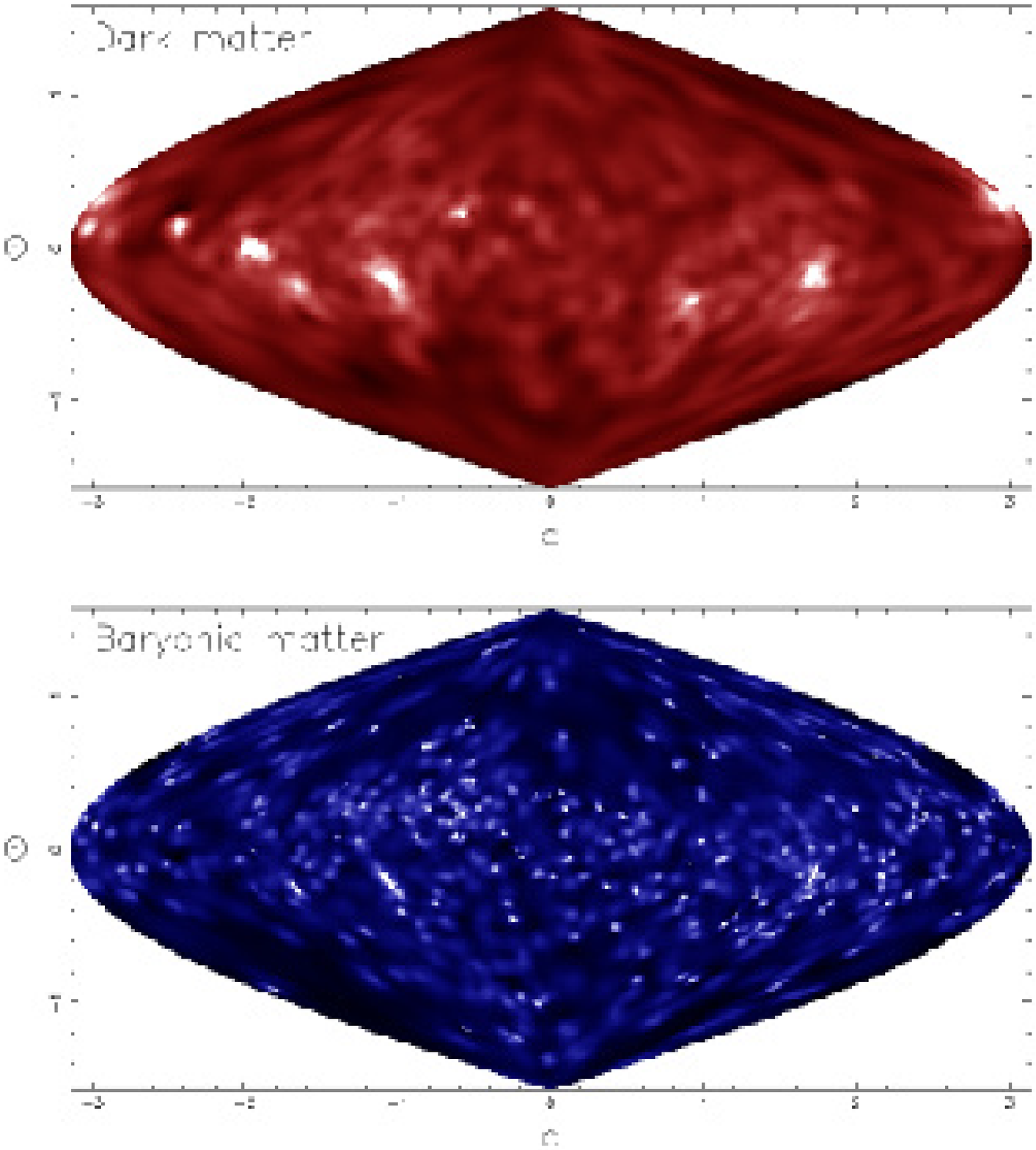}} \\  
\end{tabular}  
\caption{Galactocentric accretion rate maps for galaxies 2,3,8 and 9 (left to   
right and  
top to bottom). Light is for high accretion, dark for low accretion.   
The maps are  
integrated between $z=2$ and $z=0$, with a polar axis reoriented  
to follow the angular momentum of the baryonic matter within 50 kpc of the   
galactic center. Accretion is detected in a 50 kpc to  
100 kpc shell centered on the galaxy.  
\label{acc_map}}  
\end{figure*}

\begin{figure*}[t]  
\begin{center}
\begin{tabular}{cc}  
\resizebox{!}{!}{\includegraphics{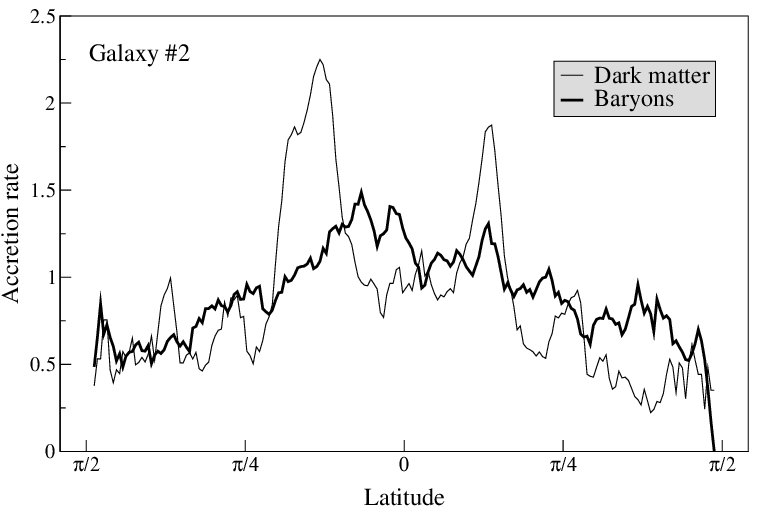}} &  
\resizebox{!}{!}{\includegraphics{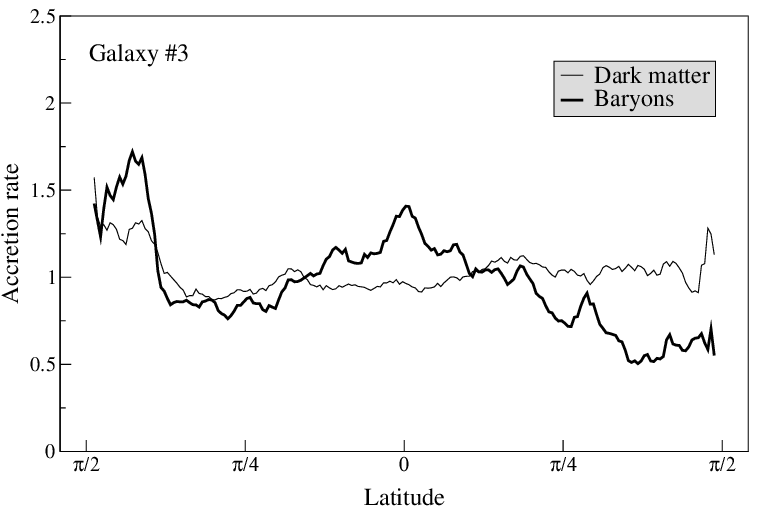}} \\ 

\resizebox{!}{!}{\includegraphics{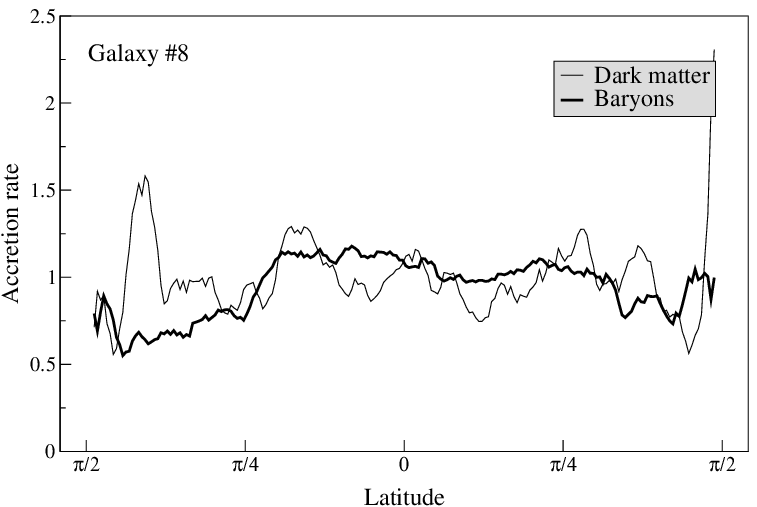}} &  
\resizebox{!}{!}{\includegraphics{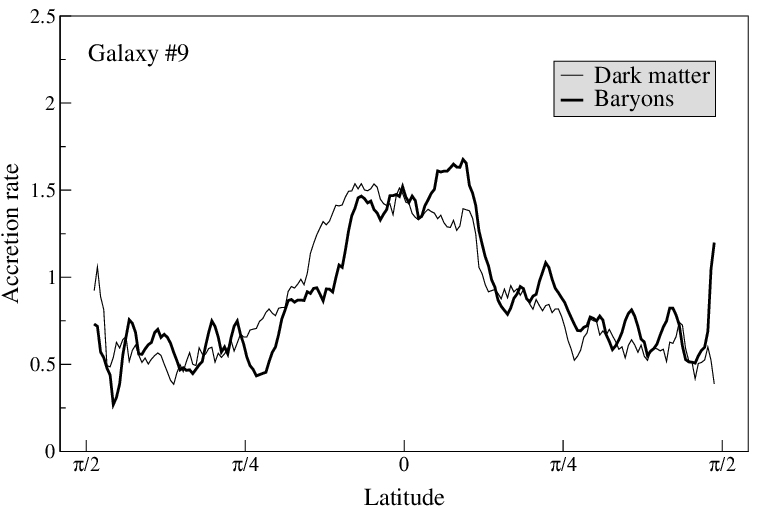}} \\ 
\end{tabular}
\caption{ Accretion rates as a function of galactic latitude for galaxies 2,3,8
and 9 (left to right and top to bottom). These maps are derived from Figure \ref
{acc_map}. Mergers have been removed to a large extent by
cutting off maps in Fig \ref{acc_map}  at 2.5 the average value of the map, then
integrating along the longitude.}  
\label{acc_map_lat}
\end{center}
\end{figure*}

The geometry of accretion is an important issue. Is the accretion isotropic?  
Is it confined to the galactic plane or does it come from one or several   
intergalactic filaments in specific directions? To answer these questions we have  
plotted angular accretion maps.  
The galaxies are identified at $z=2$. In  
subsequent outputs, those particles with a distance to the galactic center   
between 50 $h^{-1}$ kpc and 100 $h^{-1}$ kpc that have not been tagged as   
accreting particles in a previous output and that  
are within 50 $h^{-1}$ kpc of the galatic center in the last considered   
output (at $z=0$) are   
included  in the accretion map: we record their angular position on a sphere  
centered on the galaxy. For each scanned output we reorientate the pole axis  
of our angular coordinates system with a single rotation from the previous  
orientation, to keep it parallel to the spin of the baryonic matter in the  
galaxy. The spin direction can change by 90 degrees over the life time of the galaxy.

Fig. \ref{acc_map} shows angular accretion maps for  
baryonic matter and dark matter for galaxies 2, 3, 8 and 9. We have computed  
the maps for all galaxies. Removing maps with poor definition or  
major mergers, we selected these four as typical cases showing the variety  
of accretion geometry.  
They reveal that accretion is neither isotropic nor always closely   
confined in the galactic plane. The accretion is very   
anisotropic and clumpy for baryonic matter, and somewhat   
smoother if not isotropic for dark matter. This reflects the global clustering  
properties of baryonic and dark matter.  
  
These maps, integrated over  
$\sim 10$ Gyr, contain a few $10^4$ particles. It would be interesting to  
have {\sl instantaneous} accretion maps, integrated over much shorter periods  
(of the order of a galactic rotation). However, at the current mass   
resolution of the simulations, these maps would be very poorly sampled.  
Studying the evolution of the accretion maps with time is very important, but  
will have to wait for higher resolution simulations.  
  
To answer the simple question of whether accretion occurs mainly in the galatic  
plane or not, we have integrated the angular map along the longitude to plot  
the accretion rate as a function of galatic latitude. Minor mergers were  
removed before integration by applying a cut off at 2.5 times the average value  of
the angular maps. Thus the accretion rates as function of latitude shown in  
Fig. \ref{acc_map_lat} pertain to smooth accretion, not to merging events. The  
first observation is that galaxies 2, 3 and 9 baryonic matter accretion rates  
are stronger in the galactic plane than at high latitude by a factor of up to $\sim 3$.  
 This is not the case for galaxy 8, for which the galactic plane is  
not favored. The dark matter accretion rate seems to be flatter  
(less latitude dependent) than the baryonic matter accretion rate: this is  
obvious for galaxy 2, and also true for galaxy 1 if we neglect the two peaks   
linked to not-fully-removed mergers. Accretion rates for both types of matter  
are rather flat for galaxy 8, and peaked at low latitude for galaxy 9.  
  
Thus, accretion often seems to be stronger in the galactic plane for   
baryonic matter. While this also happens for dark matter, the anisotropy seems  
weaker. We have studied a few examples at high resolution; a statistical  
study at the same resolution is highly desirable. Then, connecting the  
degree of anisotropy to the environment would be possible.  

\section{Conclusion}  
  
Our goal in this work was to describe the detailed properties of   
accretion of baryonic and dark matter onto well-resolved galaxies in a cosmological  
 framework. The competition between mergers and  
smooth accretion during the mass growth of galaxies depends on the local  
cosmological environment and determines the morphological evolution of  
the galaxy. Murali {\sl et al} (\cite{Murali}) find that the mass gained by  
 accretion dominates the mass gained through mergers by a factor of 2 to 4,  
on average. This factor is much larger for field galaxies. Consequently,   
characterizing accretion precisely is necessary. However studying accretion  
in a cosmological context requires us to span a scale range from 1 kpc to sereval  
tens of Mpc. To achieve this scale range, we have developed a new numerical  
approach; the {\sl multizoom method}.  
  
The key idea derives from traditional multiresolution techniques like   
AMR or the treecode + single zoom technique. We compute a series of  
successive simulations in smaller and smaller nested boxes at higher and higher   
resolution. At each level, we record the matter inflow for the next simulation  
box and tidal field on a grid for this box. We use the data  
recorded at the previous level to compute the contribution to the dynamics of  
the matter located outside the current simulation box. We do not take into  
account the feedback that the new structures created by the increase in  
mass resolution can have on external structures and their tidal action.  
To check that this feedback is negligeable we ran a test simulation: the  
collapse of a self-gravitating pancake. We ran one high resolution simulation  
($64^3$ particles),  
one low resolution simulation ($32^3$ particles) and one zoom simulation with   
inflow and tidal field recorded from the low resolution simulation. The  
zoom and high resolution simulations have the same mass resolution. We find  
that they give very similar results, away from the  
zoom simulation box boundaries. We present velocity profiles and   
velocity dispersion profiles of the hydrodynamical shocks which form during  
the collapse. The match between the high resolution and the zoom simulation  
is satisfactory.  
  
The main advantage of the multizoom technique is the CPU cost: the high   
resolution dynamics are only computed in a small fraction of the initial  
simulation box where the galaxy is located. We were able to run simulations  
with an equivalent resolution of $256^3$ particles in 200 to 300 single  
CPU hours for each investigated region. Disk space is required to store the   
tidal field data,  
but it is not prohibitive. The method  
is implemented for parallel computers using OpenMP.  
  
We have applied the multizoom method with 4 zoom levels (20, 10, 5 and 2.5 h$^{-1}$ Mpc  
box size) and a multiphase Tree-SPH code to study the properties  
of accretion during galaxy formation in the cosmological environment. Using  
a friend-of-friend algorithm we have identified 10 galaxies with at least  
5000 baryonic matter particles at $z=0$. These galaxies range in mass from  
$3.7 10^{10}$M$_{\odot}$ to $1.9 10^{12}$M$_{\odot}$ and from $2.2 \%$ to   
$24.4 \%$ in gas fraction of the baryonic mass. We plotted a Tully-Fisher   
relation  
and found a correct slope with a shifted zero-point (similar to Navarro and  
 Steinmetz \cite{Navarro00}). We assume that this issue does not affect the properties of  
accretion at radii larger than 50 kpc, which is the focus of this work.  
  
We estimate the contribution of mergers with a mass ratio larger than   
$1 \over 20$ to the growth of the galactic mass for $z < 2$. We find that it   
is dominated by smooth accretion.  
Then we compute the evolution of the baryonic mass accretion  
rate with a 125 Myr time resolution, excluding merger events. We find   
accretion rates between 1 and 100  
M$_{\odot}$/yr almost exclusively in the form of cold gas. These rates are in   
the same range as global star formation rates, suggesting that the   
accretion regulates the star formation rate by replenishing the gaseous phase   
in the galactic disks. Low mass galaxies show  
accretion rates with an exponential decay law with a typical  time-scale  
of a few Gyr. Larger galaxies have sustained accretion rates thoughout their  
lifetime. It is possible that numerical overmerging depletes the accretion   
reservoir of small galaxies early on, leading to fast decaying accretion.   
For large galaxies, inadequate modeling of feedback from AGN or   
ram-pressure stripping from the intracluster medium may   
lead to overestimated accretion rates at late times.  
  
Then we plotted angular accretion maps for 4 typical galaxies. These maps are   
integrated between $z=2$ and $z=0$, which gives a few $10^4$ particles per   
maps. They reveal anisotropic accretion, especially for the baryonic matter  
which accretes in the form of small clumps. To check systematic anisotropy  
we integrate along galactocentric longitude, obtaining accretion rates as   
a function of galactic latitude. We find a majority of cases with an accretion  
excess at low latitude, in the galactic plane. This is consistent with results  
from Aubert {\sl et al.} (\cite{Aubert}). The additional information is that  
the excess is larger for baryonic matter than for dark matter.  
  
We have provided a few guide-lines to prescribe accretion around an isolated  
galaxy. Larger statistic samples and higher resolution are needed to confirm and  
extend these results.

\begin{acknowledgements}  
We are grateful to Fr\'ed\'eric Bournaud and Romain Teyssier  
 for interesting discussions.  
The numerical simulations in this work have been realized on the IBM-SP4  
of the CNRS computing center, at IDRIS (Orsay, France).   
\end{acknowledgements}

\end{document}